\documentclass{svjour2}                    % onecolumn
\smartqed  % flush right qed marks, e.g. at end of proof
\usepackage{graphicx}
 \journalname{JLTP}
\begin{document}

\title{Calorimetric readout of a superconducting proximity-effect thermometer}
\subtitle{}

\author{M. Meschke         \and
        J. T. Peltonen      \and \\
        H. Courtois        \and
        J. P. Pekola        }
\institute{M. Meschke         \and
        J. T. Peltonen      \and
                J. P. Pekola  \at
              Low Temperature Laboratory, Helsinki University of Technology, P.O. Box 3500, 02015 TKK, Finland  \\
             \email{meschke@ltl.tkk.fi}           %  \\
           \and
           H. Courtois  \at
              Low Temperature Laboratory, Helsinki University of Technology, P.O. Box 3500, 02015 TKK, Finland; Institut N\'eel, CNRS and Universit\'e Joseph Fourier, 25 Avenue des Martyrs, BP 166, 38042 Grenoble, France}

\date{Received: date / Accepted: date}
\maketitle

\begin{abstract}
A proximity-effect thermometer measures the temperature dependent critical supercurrent in a long superconductor - normal metal - superconductor (SNS) Josephson junction. Typically, the transition from the superconducting to the normal state is detected by monitoring the appearance of a voltage across the junction. We describe a new approach to detect the transition based on the temperature increase in the resistive state due to Joule heating. Our method increases the sensitivity and is especially applicable for temperatures below about 300 mK.  
\end{abstract}

\section{Introduction}
\label{intro}
A wide variety of methods has been realized in the past for low temperature thermometry on the mesoscopic scale, for a recent review see \cite{Giazotto2006}. One promising approach among them is the temperature dependence of the proximity induced supercurrent in a mesoscopic SNS Josephson junction. This device can easily be adjusted to the desired working range of temperatures, has a well described response to temperature, and ideally, its dissipationless supercurrent does not cause any self heating of the system. Practically, one either detects repeatably the transition from the dissipationless state to the resistive state via the voltage drop across the device or uses an alternative readout scheme exploring the kinetic inductance of the junction \cite{Giazotto2008}. We present in this work a new detection scheme monitoring the overheating of the electrons with SIN tunnel probes once the sensor switches to the resistive state, a direct application of a recent study revealing the origin of hysteresis in proximity Josephson junctions \cite{Courtois}. At the beginning, we focus on the working principle of the sensor and discuss the thermal properties of our device in detail. In what follows, we present the experimental setup and discuss our experimental results.

\section{Proximity-effect thermometer}
\label{prox}

A SNS junction length $L$ much larger than the superconducting coherence length $\xi_{\rm{s}} = \sqrt{\hbar D/2 \Delta}$ results in a small Thouless energy $E_{\rm{T}}=\hbar D / L^2$ compared to the superconducting gap of $\Delta$. In the so-defined long junction limit, the Thouless energy sets the magnitude of the proximity effects \cite{Courtois1999} and the critical current \cite{Dubos}. The critical supercurrent ($I_{\rm{C}}$) in a long SNS Josephson junction is a potential thermometer as it depends on temperature and can be tuned by the length of the normal metal island to match the desired temperature range \cite{Zaikin}. In the high temperature limit $k_{\rm{B}} T > E_{\rm{T}}$, the critical current writes:
\begin{equation}
I_{\rm{C}}=c\frac{(k_{\rm{B}}T)^{3/2}}{eR_{\rm{N}}\sqrt{E_{\rm{T}}}}\exp(-\sqrt{\frac{2\pi k_{\rm{B}}T}{E_{\rm{T}}}})
\label{ic}
\end{equation}
with a geometry dependent prefactor c \cite{Dubos,Heikkila}. In this regime, we can assume a linear temperature dependence of $I\rm_{C}$ on a semi-logarithmic plot. Fig. \ref{heat} top left shows an exemplary curve calculated with realistic parameters of our device, including a typical reduction parameter of $\alpha$ = 0.5 which describes the non-ideality of the specific SN interfaces \cite{Courtois,Heikkila}.  At very low temperatures so that $k_B T < E_{\rm{T}}$, I$\rm_{C}$ will saturate to a value corresponding to $10.82 E_{\rm{T}}/eR_{\rm{N}}$. This regime will correspond to temperatures well below 30 mK in our case.

One advantage of the proximity-effect thermometer compared to a SIN thermometer \cite{schmidt2003} is the good sensitivity towards low temperatures where the observable supercurrent increases, facilitating its readout. On the contrary, the use of a SIN thermometer gets increasingly difficult towards low temperatures as the bias current has to be reduced in order to preserve its sensitivity. A typical bias current for the SIN probe at 50 mK is below 10 pA \cite{Giazotto2006} as the voltage drop across the junction should not exceed half the gap value to reduce any cooling or heating of the electronic system. Consequently, the impedance of the SIN junction at the working point reaches a few hundred M$\rm\Omega$ to a few G$\rm\Omega$ at low temperatures. Even good quality SIN tunnel junctions show leakage resistances on the same order of magnitude \cite{dynes,averin}, setting a lower practical temperature limit for SIN thermometry. Using lower resistance tunnel junctions would result in a significant sub-gap Andreev current, which would also forbid an accurate thermometry \cite{PRL-Sukumar2}.

\section{Thermal model}
\label{sec:2}
\begin{figure}
\begin{center}
  \includegraphics[width=1\textwidth]{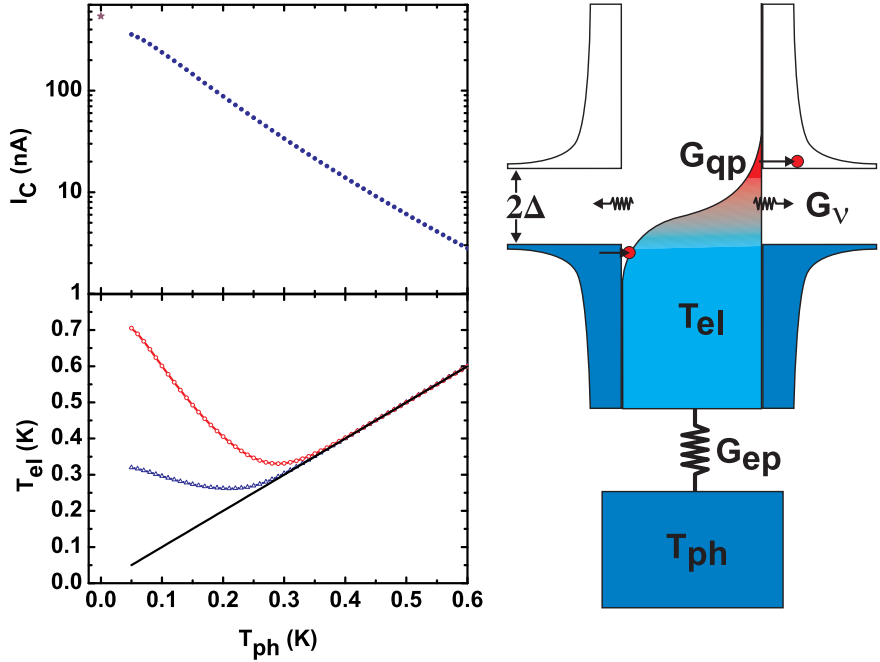}
\caption{(Color online) Thermal model of the experiment. Top left: critical current of the SNS thermometer calculated for the actual sample parameters in the high temperature regime (see text). The point at T = 0 (star) indicates the $I\rm_{C}$ value of $10.82 E_{\rm{T}}/eR_{\rm{N}}$. Right: the thermal model in detail. The electronic system in the normal metal island is in first place thermalized via electron-phonon coupling ($G\rm_{ep}$) to the lattice. Radiative thermalization ($G\rm_{\nu}$) can play a role at the low temperature end to thermalize the electronic system. Quasi-particle heat conduction ($G\rm_{qp}$) into the superconductor sets in as soon as the temperature is elevated. Bottom left: calculated temperature of the electronic system in the normal state of the SNS thermometer. The red line with open circles shows the elevated temperature assuming only electron-phonon coupling and a heat input of $R_{\rm{N}}I_{\rm{C}}^2$. Blue triangles show the same situation taking heat transport through the superconductor into account.}
\label{heat}
\end{center}
\end{figure}

The relevant process for electron thermalization in our setup (see Fig. \ref{heat} right) depends on temperature: electron-phonon coupling dominates for the temperature range from few tens of milli-kelvin up to about 300 mK. For higher temperatures, heat is transported increasingly by quasi-particles through the superconducting leads. On the contrary, the heat conductivity through the tunnel junctions with an electrical resistance of about 50 k$\rm\Omega$ is negligible with respect to the conductivity through the clean contacts. We do not include photonic heat transport \cite{schmidt,meschke} into the thermal model, although it can play a role at the low temperature end. The crossover temperature between dominating photonic heat transport and electron-phonon coupling is in the present setup strongly reduced as the island resistance of about 10 $\rm\Omega$ is typically one to two orders of magnitude too low for a good matching to the environment.

At the clean contacts between the island and the superconducting electrodes, the latter guarantee a good thermal isolation below temperatures of approximately 0.3 $T\rm{_C}$,\cite{bardeen} where $T\rm{_C}$ is the superconducting critical temperature. The thermal conductivity of the superconductor with respect to the normal state conductivity ($L_{\rm{0}}g_{\rm{N}}T$, with the Lorentz number $L_0=2.45\times10^{-8}$ $\rm{W\Omega/K}^2$ and the normal state electrical conductivity $g_{N}$) decays in the low temperature limit as ${\rm{exp}}(-\Delta/k_{\rm{B}}T)$. It is difficult to derive an exact number for $g_{\rm{N}}$ as the sample design does not allow a direct measurement of the normal state resistance between the island and the copper shadow below the superconducting electrode. Instead we show a calculation in Fig. \ref{heat} with a realistic value $g_{\rm{N}} =$ 3 $\rm \Omega$. Eventually, we can not exclude an enhanced heat transport through the superconductor to the normal metal shadows due to the inverse proximity effect \cite{belzig,EPL-Sillanpaa}.

We consider that the thermalization of the electronic system in our normal metal island with a volume $\mathcal{V}$ is governed by the electron-phonon coupling,
\begin{equation}
P=\Sigma \mathcal{V} (T _{\rm{{el}}}^5 - T _{\rm{ph}}^5)
\label{ep}
\end{equation}
where $\Sigma$ is determined by the mass density and the longitudinal sound velocity of the material  \cite{Hekking}. Experimentally, one finds values of 2$\times$10$^9$ \rm{Wm}$^{-3}$\rm{K}$^{-5}$ for copper \cite{Roukes,Clarke}. We neglect here the influence of the proximity effect and assume that the resulting modification of the density of states in the normal metal has only marginal influence on the electron-phonon coupling strength.

A simple estimation clarifies the working principle of the proposed method: the resulting electronic temperature in the normal state of the thermometer is determined by the corresponding Joule heating of the order of $R_{\rm{N}}I\rm_{C}^2$. The transition from the superconducting state at $T$ = 50 mK at a critical current of $I{\rm_{C}}$= 300 nA results in a power of about 1 pW. This power would increase the temperature of the electronic system in the normal metal island according to Eq. (\ref{ep}) to about 700 mK. The quasiparticle heat conduction will reduce this value, but the remaining temperature increase of few hundred milliKelvin is still easily detectable with the SIN probe. Figure \ref{heat} depicts the resulting normal state electron temperatures of the island as a function of the base temperature, assuming electron-phonon coupling and also quasiparticle heat conduction of the island at current $I_{\rm{C}}$.

Consequently, the calorimetric readout will improve due to two effects towards lower temperatures: both decreasing electron-phonon coupling and increasing critical current of the proximity-effect thermometer amplify the detectable increase of the electronic temperature in the normal state of the SNS junction significantly. An upper useful limit on the order of 300 mK for the described method is given by the diminishing heat power due to the smaller critical current and the enhanced electron thermalization due to the stronger electron-phonon coupling and the increased quasiparticle heat conduction.

\section{Experimental setup}
\begin{figure}
\begin{center}
\includegraphics[width=0.8\textwidth]{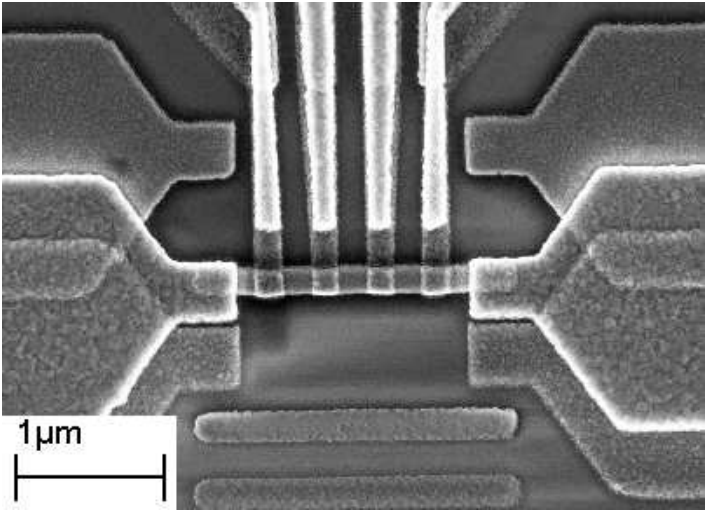}
\caption{Scanning electron microscope image of the proximity-effect thermometer. Three shadows are visible from the different metalization steps. They are (from bottom to top): 30 nm thick aluminum with oxidation to form the tunnel junctions, 70 nm aluminum for the clean contacts to the normal metal island, 27 nm copper for the normal metal island. As a result, four tunnel barriers (along the island) and two clean interfaces (at both ends) connect the copper island to superconducting aluminum leads. The normal metal island has dimensions of 1.6 $\times$ 0.15 $\mu$m. Note that the thick superconducting leads are also in direct contact with the underlying normal metal at a distance of approximately 1 $\mu$m. The contrast between copper and aluminum is weak as the image was taken with a surface sensitive in-lens-detector to illustrate mainly the outlines of the structures.}
\label{fig:1}
\end{center}
\end{figure}

The sample (see Fig. \ref{fig:1}) is fabricated using standard e-beam lithography with a two layer, approximately 1.8 $\mu$m thick PMMA resist and three angle shadow evaporation. The central element is the normal metal island made out of a 27 nm thick copper film (square resistance 1.2 $\rm\Omega$). The island is placed in between two 70 nm thick superconducting electrodes with a clean contact overlap of 250 nm to realize the long Josephson junction ($L$ = 1.6 $\mu$m). These electrodes have an increased thickness to reduce the inverse proximity effect and to improve consequently the definition of the junction length. The diffusion coefficient $D$ in Cu is determined by the measured normal state resistance to be 110 cm$^2$/s for our sample. It gives a calculated Thouless energy $E_{\rm{T}} \approx$ 2 $\rm\mu eV$ well below the superconducting gap $\Delta \approx$ 200 $\mu eV$. Four tunnel junctions are formed between oxidized aluminum probes and the copper island with a resulting tunnel resistance of about 50 k$\rm\Omega$ each. For a good junction quality and reproducibility, it is essential to minimize overheating of the sample during the e-beam evaporation to avoid contamination of the metals with outgassing components of the resist. A long distance between the sample and the evaporation source, a relatively small metal target ($\approx$ 1 cm in diameter) and a well focused electron beam are helpful to keep the substrate temperature, measured in situ with a pyrometer, below 50 $^{\rm o}$C during metalization.

A dilution refrigerator with a base temperature below 50 mK is employed in all measurements. The DC lines in the cryostat are filtered via one meter long thermocoax lines and equipped with cold resistors ($R \approx$ 300 $\rm\Omega$) on the sample stage. Generally, all current and voltage sources and the amplifiers are at room temperature.

\begin{figure}
\begin{center}
\includegraphics[width=.7\textwidth]{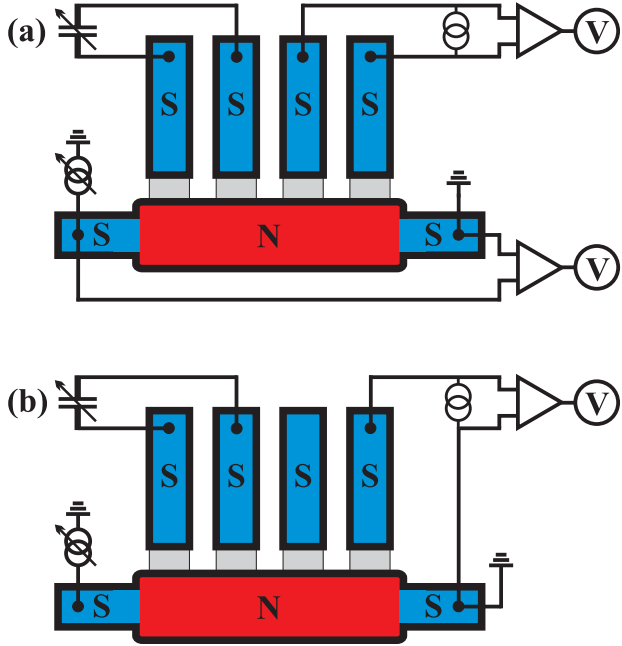}
\caption{(Color online) Schematic drawing of the sample together with the setups for the calorimetric readout of the Josephson thermometer. (a) The two clean contacts at both ends of the normal metal island (N) to the superconducting leads (S) are connected to two pairs of DC lines to allow the necessary four-probe measurement of the (SNS) Josephson thermometer. Each of the four tunnel probes is connected with a single DC line. One pair of junctions is voltage biased to allow cooling or heating of the electronic system in the normal metal island. A current bias of typically 6 pA applied to the second pair of tunnel junctions is used for thermometry. (b) Alternative setup that is used in the present work: a single tunnel junction is current biased to allow for thermometry. The voltage amplifier across the SNS junction can be omitted.}
\label{fig:2}
\end{center}
\end{figure}

The experimental setup in Fig. \ref{fig:2} (a) allows a full characterization of the sample: the SNS thermometer is current biased with a bipolar voltage sweep in combination with a 1 M$\rm\Omega$ resistor at room temperature. The voltage drop across the device ($R_{\rm{N}}I\rm_{C}$) is of the order of a few $\mu$V, which is small compared to the voltage drop along the lines of the cryostat (of resistance of about 150 $\rm\Omega$) and has to be amplified by a gain of 10$^6$. Consequently, a four probe configuration is indispensable for the measurement. A floating voltage source biasing one pair of tunnel junctions with a voltage sweep in the range $\pm 3 \Delta$ allows cooling and heating of the electronic system \cite{Leivo}. Finally, the remaining two tunnel junctions are biased with a high impedance (20 G$\rm\Omega$) constant current source of 6 pA to allow for thermometry. This thermometer probe has a voltage response of the order of half the superconducting gap value (0.2 mV) and a high impedance (in the range of few tens of M$\rm\Omega$). Consequently, it can be measured in a two probe configuration with moderate amplification (10$^4$).

The lower panel of Fig. \ref{fig:2} (b) depicts an alternative setup using exclusively the calorimetric signal for the readout of the SNS thermometer: the voltage probe along the normal metal island is replaced by a single current biased SIN tunnel junction, probed with respect to the normal metal island.

\section{Experimental performance of the thermometers}

\begin{figure}
\begin{center}
\includegraphics[width=1\textwidth]{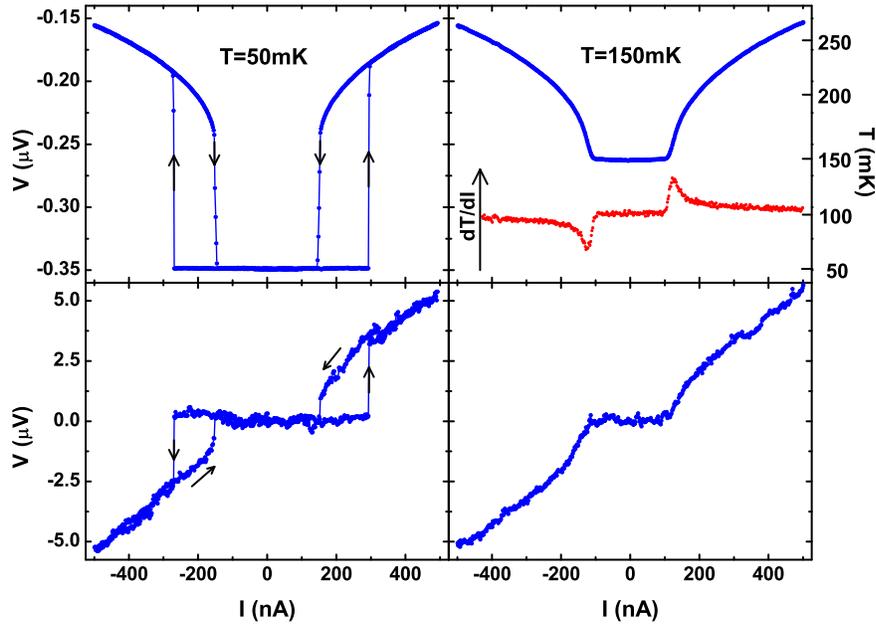}
\caption{(Color online) Experimental comparison between the calorimetric readout (top) of the SNS thermometer and the standard method (bottom) at bath temperatures of 50 mK and 150 mK. The arrows indicate the direction of the current sweep for the low temperature case, where hysteresis due to overheating is present. In the top right panel, the temperature scale arises from a calibration of the SIN probe and the red dots show the derivative of the electronic temperature.}
\label{2t}
\end{center}
\end{figure}

Figure \ref{2t} shows a comparison between the measured voltage drop across the SNS and the SIN thermometer readout at two different base temperatures obtained with the setup of Fig. \ref{fig:2} (a). At $T$ = 50 mK, both voltages indicate clearly the transition of the SNS junction to the normal state once the critical current is exceeded. Obviously, the signal to noise ratio of the SIN probe indicating the temperature increase is clearly superior over that from the direct voltage measurement across the SNS junction. At an elevated temperature, the advantage of the SIN probe is even more pronounced: a clear temperature increase is still observed whereas the starting point of the voltage drop across the SNS junction is difficult to distinguish due to the noise level. The derivative of the electronic temperature indicates that a sharp indication for the transition point still exists.

\begin{figure}
\begin{center}
\includegraphics[width=1\textwidth]{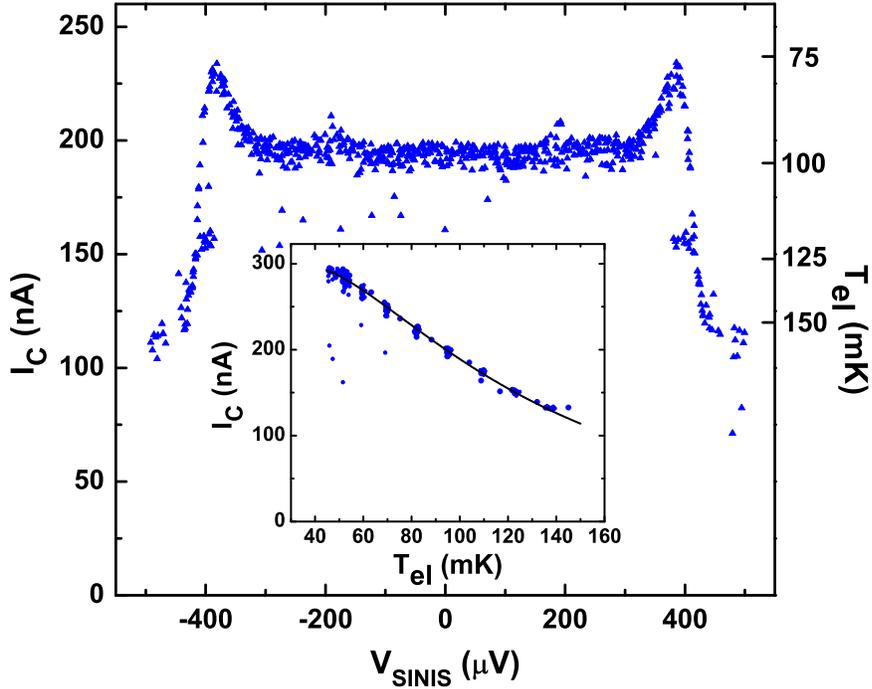}
\caption{(Color online) Electronic temperature measured with the calorimetric readout of the SNS thermometer as a function of the bias voltage $V_{\rm{SINIS}}$ across the SINIS cooler. The inset shows the calibration of the SNS thermometer versus the electronic temperature: the solid line is a fit to the data points based on Eq. (\ref{ic}).}
\label{cool}
\end{center}
\end{figure}

Finally, an application of the calorimetric SNS thermometry using the setup in Fig. \ref{fig:2} (b) is depicted in Fig. \ref{cool}: the electronic temperature of the normal metal island is cooled down from the base temperature of 100 mK to about 75 mK when the SINIS cooler junctions are biased at  2$\rm\Delta \approx$ 400 $\mu eV$. The temperature scale at the right axis of the main panel is generated using the data points of the inset. Note that the electronic temperature of the island does not saturate even towards the lowest measured temperature below 50 mK, contrary to other experiments \cite{meschke,Juha} where the island is connected to an identical environment exclusively via tunnel junctions. This improved thermalization may be attributed to the described photonic heat transport. Alternatively, one can assume an enhanced heat transport through the superconductor to the normal metal shadows due to the inverse proximity effect.

\section{Conclusion}

We present a new calorimetric readout of a proximity-effect thermometer. The main advantages lie in the better sensitivity to the transition point from the superconducting to the normal state of the thermometer combined with a simplified experimental setup. The method also reveals a detailed insight into the underlying thermal processes and is straightforward to implement, especially if tunnel junctions are already fabricated for cooling purposes. Designing a thermometer for a temperature range spanning from 10 mK up to few 100 mK will benefit from the proposed method, as the detection of the small critical current at the high temperature end benefits significantly from the improved sensitivity.

\begin{acknowledgements}
We acknowledge the financial support from from the ANR contract "Elec-EPR", the ULTI-3 and NanoSciERA "Nanofridge" EU projects. H. Courtois thanks the Low Temperature Laboratory for hospitality.
\end{acknowledgements}


\end{document}